\documentclass[11pt]{article}
\usepackage{times}

\usepackage{fullpage}
\usepackage[bf,small]{caption}

\usepackage{amssymb,amsmath}
\usepackage{graphicx}
\usepackage{url}
\makeatletter
\def\url@leostyle{%
  \@ifundefined{selectfont}{\def\UrlFont{\small\sffamily}}%
  {\def\UrlFont{\small\sffamily}}%
}
\makeatother	
\usepackage{tikz}
\usepackage{subfigure}
\usepackage{indentfirst}

\newcommand{\descr}[1]{\vspace{0.25cm} \noindent \textbf{#1}}

\def\naive{na\"{i}ve}

\newcommand\textsubscript[1]{\@textsubscript{\selectfont#1}}
\def\@textsubscript#1{{\m@th\ensuremath{_{\mbox{\fontsize\sf@size\z@#1}}}}}
\newcommand\textbothscript[2]{%
  \@textbothscript{\selectfont#1}{\selectfont#2}}
\def\@textbothscript#1#2{%
  {\m@th\ensuremath{%
    ^{\mbox{\fontsize\sf@size\z@#1}}%
    _{\mbox{\fontsize\sf@size\z@#2}}}}}

\definecolor{darkblue}{RGB}{47,79,79}

\usepackage[bookmarks=false, colorlinks=true, plainpages = false,  linkcolor = darkblue,   citecolor = darkblue, urlcolor = darkblue, filecolor = blue]{hyperref}

\newcommand{\Q}{\ensuremath{\mathcal{Q}}}
\newcommand{\M}{\ensuremath{\mathcal{M}}}

\newcommand{\remove}[1]{}

\newcommand{\PS}{\textsf{\small PS}}

\begin{document}

\title{\bf Participatory Privacy:\\ Enabling Privacy in Participatory Sensing
}

\author{
Emiliano De Cristofaro \\
\normalsize Palo Alto Research Center (PARC) \\
\normalsize edc@parc.com \and
Claudio Soriente\\
\normalsize ETH Zurich, Switzerland\\
\normalsize claudio.soriente@inf.ethz.ch
}

\date{}

\maketitle

\begin{abstract}
Participatory Sensing is an emerging computing paradigm that enables the
distributed collection of data by self-selected participants.
It allows the increasing number of  mobile phone users to share local knowledge
acquired by their sensor-equipped devices, e.g., to monitor  temperature, pollution level or
consumer pricing information.
While research initiatives and prototypes proliferate, their real-world impact
is often bounded to comprehensive user participation.
If users have no incentive, or feel that their privacy might be endangered,
it is likely that they will not participate.

In this article, we focus on privacy protection in Participatory Sensing and introduce a suitable
privacy-enhanced infrastructure.  First, we provide a set of definitions of privacy requirements for
both data producers (i.e., users providing sensed information) and consumers (i.e., applications accessing the data).
Then, we propose an efficient solution designed for mobile phone users, which incurs very low overhead.
Finally, we discuss a number of open problems and possible research directions.
\end{abstract}

\section{Introduction}\label{sec:introduction}
In the last decade, researchers have envisioned the outbreak of Wireless Sensor Networks (WSNs)
and predicted the widespread installation of sensors, e.g., in infrastructures, buildings, woods, rivers, or even the atmosphere.
This has triggered a lot of interest in many different WSN topics, including identifying and addressing security issues, such as
data integrity, node capture, secure routing, etc. On the contrary, privacy has not really been a concern in WSNs,
as sensors are usually owned, operated, and queried by the same entity. (For instance,
the National Department of Transportation deploys sensors and collects traffic information related to national highways.)

On the other hand, the proliferation of mobile phones, along with their pervasive connectivity,
has propelled the amount of digital data produced and processed everyday.
This has driven researchers and IT professionals to discuss and develop a novel sensing paradigm,
where sensors are not deployed in specific locations, but are \emph{carried} around by people.
Today, many different sensors are already deployed in our mobile phones, and
soon all our gadgets (e.g., even our clothes or cars)  will embed a multitude of sensors (e.g., GPS, digital imagers, accelerometers, etc.).
As a result, data collected by sensor-equipped devices becomes of extreme interest to other users and applications.
For instance, mobile phones may report (in real-time) temperature or noise level; 
similarly, cars may inform on traffic conditions. 

This paradigm is called \emph{Participatory Sensing} (\PS) --
sometimes also referred to as {\em opportunistic} or {\em urban} sensing \cite{cuff08comm}.
It combines the ubiquity of personal devices with sensing capabilities typical of WSN.
As the number of mobile phone subscriptions exceeds 5 billions, \PS\
becomes a cutting-edge and effective distributed-computing (as well as business) model.
We argue that \PS\ appreciably expands the capabilities of WSN applications, e.g., allowing effective monitoring
in scenarios where the set up of a WSN is either not economical or not feasible.

However, its success is strongly related to the number of users actually willing to commit personal device resources to
sensing applications, and thus, to associated privacy concerns.
Observe that sensing devices are no longer ``dull'' gadgets, owned by the entity querying them.
They are personal devices that follow users at all times, and their reports often expose personal and sensitive information.
Consider, for instance, a \PS\ application like {\url{http://www.gasbuddy.com/} where gas prices are monitored via user reports,
and information announced by participants inevitably exposes their current and past locations, hence, their movements.
If users have no incentive in contributing sensed data or feel that their privacy might be violated, they
will (most likely) refuse to participate. Thus, not only traditional security but also privacy issues must
be taken into account.

In this article, we focus on privacy protection in \PS.
We define privacy in this new context, present a privacy-enhanced \PS\ infrastructure,
and elaborate on a number of desirable features which constitute challenging research problems.
Proposed privacy-protecting layer can be easily adopted by available \PS\ applications to
enforce privacy and enhance user participation.

\section{Participatory Sensing}

\descr{What is Participatory Sensing?} \PS\ is an emerging paradigm that focuses
on the seamless collection of information from a large number of connected, always-on, always-carried devices, such as mobile phones.
\PS\ leverages the wide proliferation of commodity sensor-equipped devices and the ubiquity of broadband network infrastructure to
provide sensing applications where deployment of a WSN infrastructure is not economical or not feasible.
%
%
\PS\ provides fine-grained monitoring of environmental trends without the need to set up a sensing infrastructure.
Our mobile phones \emph{are} the sensing infrastructure and the number and variety of applications are potentially unlimited.
Users can monitor gas prices (\url{http://www.gasbuddy.com/}), traffic information (\url{http://www.waze.com/}), available parking spots (\url{http://spotswitch.com/}),
just to cite a few. We refer readers to~\cite{projectpage} for an updated list of papers and projects related to \PS.

\descr{What {\em isn't} Participatory Sensing?} \PS\ {is not} a mere evolution of WSN,
where motes are replaced by mobile phones.
Sensors are now relatively powerful devices, such as mobile phones, with much greater resources than WSN motes.
Their batteries can be easily recharged and production cost constraints are not as tight.
They are extremely {\em mobile}, as they leverage the ambulation of their carriers.
Moreover, in traditional WSNs, the network operator is always assumed to manage and own the sensors.
On the contrary, this assumption does not fit most \PS\ scenarios, where mobile devices are {\em tasked} to participate
into gathering and sharing local knowledge. Hence, a sensor (or its owner) might choose whether to participate or not.
As a result, in \PS\ applications, different entities co-exist and might not trust each other.

\descr{Participatory Sensing Components.} A typical \PS\ infrastructure involves (at least) the following parties:
\begin{enumerate}
\item \textbf{Mobile Nodes} are the union of a carrier (i.e., a user)  with a sensor installed on a mobile phone or other portable, wireless-enabled device.
They provide reports and form the basis of any \PS\ application.
\item \textbf{Queriers} subscribe to information collected in a \PS\ application (e.g., ``temperature in Irvine, CA'') and obtain corresponding reports.
\item \textbf{Network Operators} manage the network used to collect and deliver sensor measurements , e.g., they maintain
GSM and/or 3G/4G networks.
\item \textbf{Service Providers} act as intermediaries between Queriers and Mobile Nodes, in order to deliver report of interest to Queriers.
\end{enumerate}


Queriers can subscribe to the appropriate Service Provider for one or more type of measurements.
For example, assume that Alice subscribes to ``available parking spots on W 16th Street, New York'', or Bob is interested in the ``temperature in Central Park, New York''.
In turn, Mobile Nodes share local knowledge---either voluntary or in return for some profit---with
one or more Service Providers, that make information available to Queriers.
For example, assume Carol' mobile phone sends report ``3 available parking spots on E 56th, New York'', while
John's device sends ``$74^o F$ in Central Park, New York''.

As Mobile Nodes and Queriers have no direct communication nor mutual knowledge, Service Providers route reports matching
specific subscriptions to their original Queriers.
In fact, Mobile Nodes ignore which Queriers (if any) are interested in their reports.
For example, the Service Provider forwards John's temperature report to Bob; Carol's parking report is not sent to Alice as it
refers to a different location.

\section{Privacy Concerns}
\PS\ provides an effective solution to a wide range of applications, however, it prompts
several security and  privacy concerns that need to be carefully addressed.

On the one hand, issues such as confidentiality or integrity can be mitigated
using state-of-the-art techniques. For instance, all parties can
be protected from external eavesdroppers using SSL/TLS. The latter provides a secure channel between any two parties,
so that communications between Mobile Nodes and Service Providers or between Service Providers and Queriers are kept confidential.

On the other hand, the need for privacy protection stems from the potential leakage of personal information to \emph{internal adversaries}.
Indeed, as the Service Provider collects \emph{all} data (i.e., reports and queries),
it might learn a considerable amount of sensitive information about both Mobile Nodes and Queriers, and violate
the privacy of their movements, interests, habits, and more.
For instance, the Service Provider learns that both Bob and John are located in Central Park, New York.
It also learns that Alice is driving on W 16th Street, looking for parking.
The continuous collection of information over long periods allows the Service Provider to
meticulously profile users.

Further, as data collected through \PS\ applications becomes available to external entities
and organizations (i.e, the Queriers), query interests also become sensitive and need to be hidden.
For instance, Service Providers should not learn which interests are ``hot''.

Finally, there is a tension between privacy and accountability as \PS\ business models may require, at the very least,
that reports are available only to entitled (e.g., authorized or paying) members.

However, we claim there is one main reason to protect privacy. If users feel that their privacy is endangered,
they will deny sharing their reports. 
Specifically, it is required that the Service Provider performs report/query matching but learns no information about query interests.
Also, data reports should not reveal to the Service Provider, the Network Operator, or unauthorized Queriers, any information about
a Mobile Node's identity, its location, the type of measurement (e.g., temperature) or the quantitative information (e.g., $74^o F$).

\section{A Novel Privacy-Enhanced Participatory Sensing Infrastructure}\label{sec:pepsi}
We now present our innovative solution for a Privacy-Enhanced Participatory Sensing Infrastructure (PEPSI).
We describe its architecture and privacy desiderata, and overview our instantiation.
Finally, we discuss efficiency costs introduced by the privacy-protecting layer.

\subsection{PEPSI Architecture}
PEPSI protects privacy using efficient cryptographic tools. Similar to other cryptographic solutions, it introduces an additional (offline) entity,
namely the Registration Authority. It sets up system parameters and manages Mobile Nodes or Queriers registration.
However, the Registration Authority is not involved in real-time operations (e.g., query/report matching) nor
is it trusted to intervene for protecting participants' privacy.

\begin{figure*}[t!]
\centering
\fbox{
\includegraphics[width=.75\columnwidth]{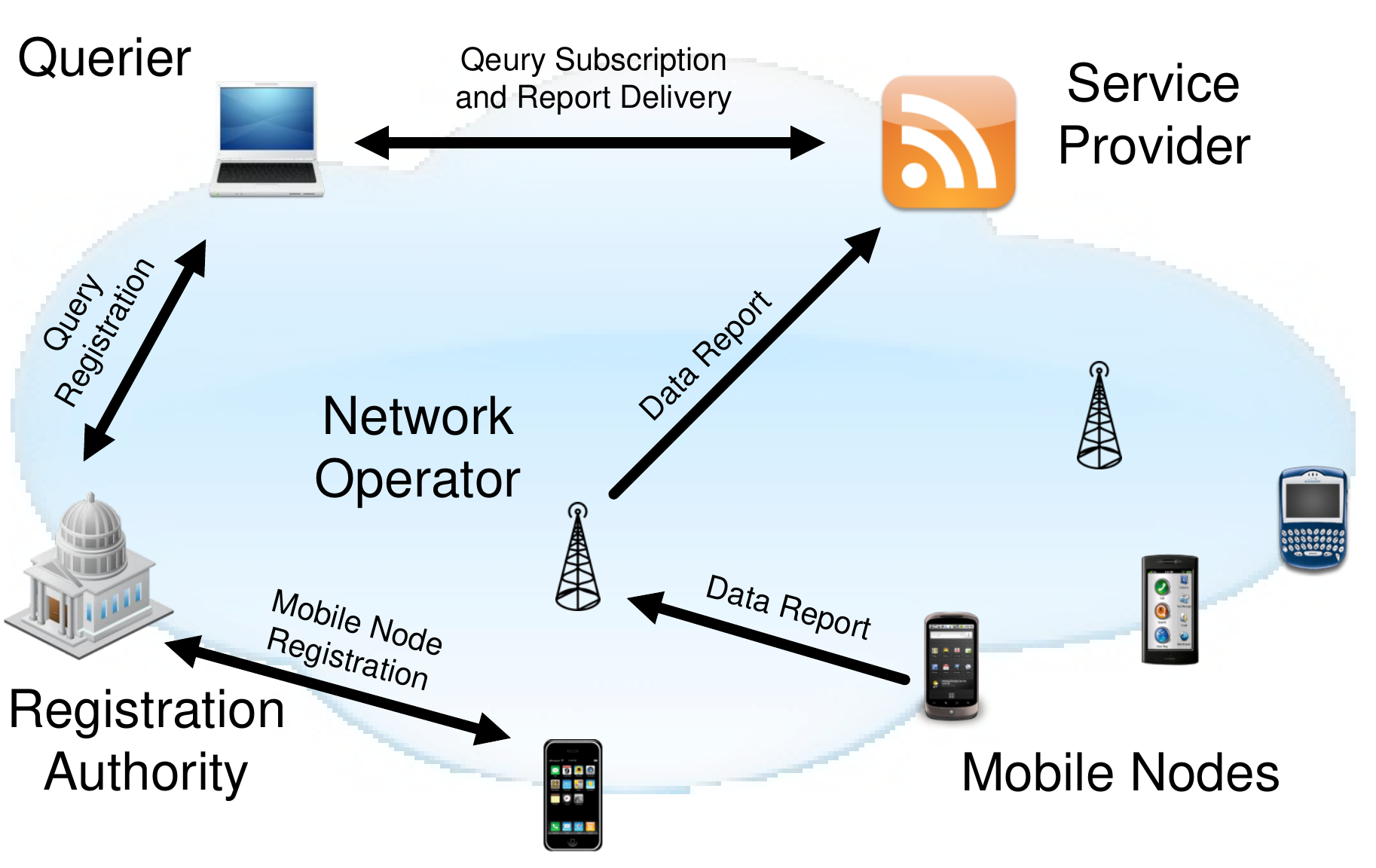}
}
\caption{Privacy-Enhanced Participatory Sensing Infrastructure.}
\vspace{0.3cm}
\label{fig:architecture}
\end{figure*}

Figure~\ref{fig:architecture} illustrates the PEPSI architecture.
The Registration Authority can be instantiated by any entity in charge of managing participants registration (e.g., a phone manufacturer).
A Service Provider offers \PS\ applications (used, for instance, to report and access pollution data)
and acts as an intermediary between Queriers and Mobile Nodes. Finally,
Mobile Nodes send measurements acquired via their sensors using the network infrastructure and Queriers are users or organizations (e.g., bikers) interested in obtaining reports (e.g., pollution levels).

PEPSI allows the Service Provider to perform report/query matching while guaranteeing the privacy of both mobile Nodes and Queriers.
It aims at providing (provable) privacy {\em by design}, and starts off with defining a clear set of privacy properties.

\subsection{Privacy Desiderata}
The {\em privacy desiderata} of \PS\ applications can be formalized as follows:
\begin{itemize}
\item \textbf{\em Soundness:} Upon subscribing to a query, Queriers in possession of the appropriate authorization
always  obtain the desired query results. 
\item \textbf{\em Node Privacy:} Neither the Network Operator, the Service Provider, nor any unauthorized Querier, learn any information
about the type of measurement or the data reported by a Mobile Node. Also, Mobile Nodes should not learn any information about other nodes' reports.
Only Queriers in possession of the corresponding authorization obtain reported measurements.
\item \textbf{\em Query Privacy:} Neither the Network Operator, the Service Provider, nor any Mobile Node or any other Querier, learn any information about Queriers' subscriptions.
\item \textbf{\em Report Unlinkability:} No entity can successfully link two or more reports
as originating from the same Mobile Node.
However, as we discuss below, we do not pursue
Report Unlinkability with respect to the Network Operator.
\item \textbf{\em Location Privacy:} No entity can learn the current location of a Mobile Node.
(Again, excluding the Network Operator).
\end{itemize}

In realistic scenarios, it appears unlikely -- if not impossible -- to guarantee Report Unlinkability and
Location Privacy with respect to the Network Operator.
In fact, \PS\ strongly relies on the increasing use of broadband 3G/4G connectivity.
In these networks, current technology does not allow to provide user anonymity with respect to the Network Operator.
Mobile Nodes are identified through their International Mobile Subscriber Identity, and any technique for
identifier obfuscation would lead to service disruption (e.g., the device would not receive incoming calls).
Further, the regular usage of cellular networks (e.g., incoming/outgoing phone calls), as well as heartbeat messages
exchanged with the network infrastructure, irremediably reveal device's location.
To provide Report Unlinkability/Location Privacy with respect to other parties, we need to trust the Network Operator
(who routes Mobile Nodes' reports to Service Providers) not to forward any information identifying the Mobile Nodes
(e.g., the identifier, the cell from which the report was originated, etc.).

\subsection{PEPSI Construction}
One of the main goals of PEPSI is to hide reports and queries to unintended parties.
Thus, those cannot be transmitted {\em in-the-clear}, but need to be encrypted.
In this section, we discuss how to achieve, at the same time, (1) secure encryption of reports and queries,
and (2) efficient and oblivious matching by the Service Provider.
Due to space limitation and to ease presentation, we only provide an overview
of our construction (with no technical details). We refer interested readers to the extended version of the paper (available
on project page~\cite{projectpage}) for
a complete description of our techniques, as well as formal cryptographic proofs.

\descr{A \naive\ solution.}
Traditional confidentiality means are not suited for \PS\ applications.
Recall that in our context, Mobile Nodes and Queriers have no mutual knowledge or common history:
that is, Mobile Nodes provide reports obliviously of (any) potential receiver, while Queriers subscribe to data reports not knowing
who (if any) will every provide measurements of interest.
Hence, we cannot assume that each Mobile Node shares a unique pairwise secret key with each Querier and
that reports are encrypted under that key via a symmetric-key cipher (e.g., AES).
Even if we were to allow interactions between Mobile Nodes and Queriers,
we would still need the former to encrypt reports under each key shared with Queriers.
This would generate a number of ciphertexts  quadratic in the number of measurements.
Alternatively, we could use a public key encryption scheme and provide Mobile Nodes with the public keys of Queriers.
Still, scalability would be an issue as each report would be encrypted under the public key of each Querier.
In general, because of scalability and loose coupling between data producers and consumers,
Mobile Nodes cannot provide measurements intended for a specific Querier and the latter cannot ask
for data from a given Mobile Node.

\bigskip
Our main building block is Identity-Based Encryption (IBE) --- a cryptographic primitive, based on bilinear map pairings,
that enables asymmetric encryption using any string (``identity'') as a public key.
In IBE, anyone can derive public keys from some unique information about the recipient's identity.
Private decryption keys are generated by a third-party, called the Private Key Generator (PKG).
\emph{Our intuition is to use a tagging mechanism on top of IBE.}

\descr{Report Encryption.}
We assume that each report or subscription is identified by a set of labels, or keywords. These are used as ``identities'' in an IBE scheme.
For example, labels ``Temperature'' and ``Central Park, NY'' can be used to
derive a unique public encryption key, associated to a secret decryption key.
Thus, Mobile Nodes can encrypt sensed data using report's labels as the (public) encryption key.
Queriers should then obtain the private decryption keys corresponding
to the labels of interest. Those are obtained, upon query registration, from the Registration Authority -- which,
in practice, acts like a PKG.

\descr{Efficient Matching using Cryptographic Tags.}
After enabling encryption/decryption
of reports, we need to allow the Service Provider to efficiently match them against queries.
In fact, the application of IBE to \PS\ settings is not trivial:
with a straightforward use of IBE, oblivious matching of queries and reports would be impossible.
In other words, the Service Provider would forward {\em all} (encrypted) reports to all Queriers;
each of them will only be able to decrypt reports of interests, i.e., the ones for which they hold the decryption keys.
However, given the large amount of reports produced by Mobile Nodes, this would incur
a considerable overhead for the Querier, that must try to decrypt all reports using each of her decryption keys.
To address this problem, we propose an efficient tagging mechanisms:
Mobile Nodes {\em tag} each report with a cryptographic
token that identifies the nature of the report only to authorized Queriers, but does not leak
any information about the report itself.
Tags are computed using the same labels used to derive encryption keys.
Similarly, Queriers compute tags for the labels defining their interests (using the corresponding
decryption keys) and provide them to the Service Provider at query subscription.

Our main contribution, in this context, is to exploit the mathematical properties of bilinear map pairings:
we ensure that, whenever a report matches a query, corresponding tags also match.
In other words, a tag computed by John using the encryption key derived from label ``temperature in Central Park, New York'',
is equal to the tag computed by Bob using the decryption key computed over the same label.
Specifically, Mobile Nodes upload reports along with the respective tags, while Queriers define their subscriptions
uploading the tags they compute at the Service Provider.
The latter can find matches (i.e., a tag related to a report equals the tag related to a subscription)
without learning any information about underlying queries/reports.

\begin{figure*}[h!]
\centering
\fbox{
\includegraphics[width=.85\columnwidth]{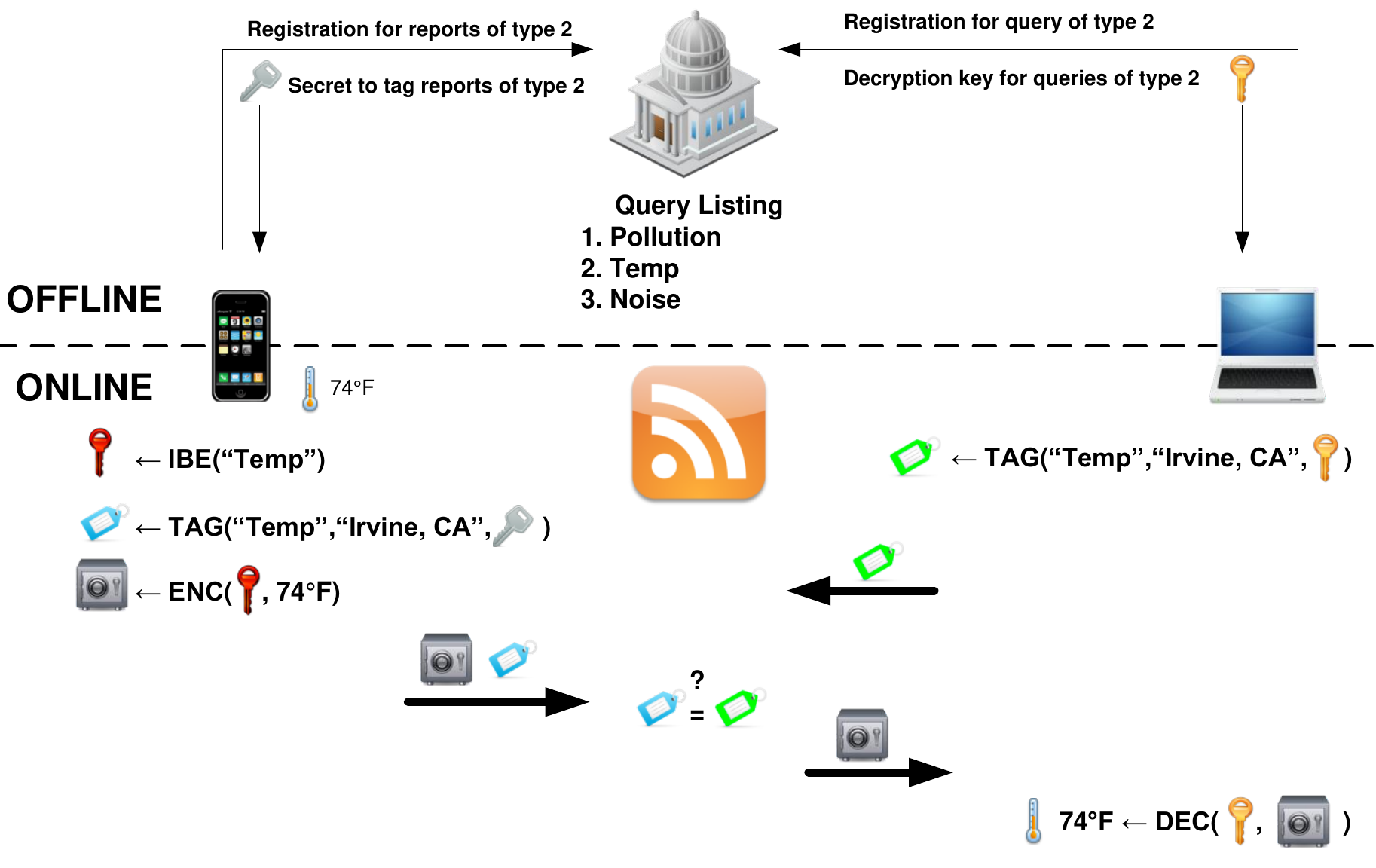}
}
\caption{PEPSI operations.}
\vspace{0.3cm}
\label{fig:operations}
\end{figure*}

\subsection{PEPSI Operations }
Figure \ref{fig:operations} shows how PEPSI work.
The upper part of the figure depicts the offline operations where the Registration Authority is involved to
register both Mobile Nodes and Queriers.

\descr{Querier Registration.}
In the example,  Querier \Q\ (the laptop on the right side)
picks ``Temp'' among the list of available queries and obtains the corresponding decryption key (yellow key).

\descr{Mobile Node Registration.}
Similarly, Mobile Node \M\ (the mobile phone on the left side) decides to report about temperature in its location and obtains the corresponding secret used
for tagging (grey key).

\medskip
The bottom part of Figure \ref{fig:operations} shows the online operations where the Service Provider is involved.
\medskip

\descr{Querier Subscription.}
\Q\ subscribes to queries of type ``Temp'' in ``Irvine, CA'' using these keywords and the decryption key acquired offline,
to compute a (green) tag; the algorithm is referred to as $\leftarrow TAG()$.
The tag leaks no information about \Q's interest and is uploaded at the Service Provider.

\descr{Data Report.}
Any time \M\ wants to report about temperature, it derives the public decryption key (red key) for reports of type ``Temp''
(via the $\leftarrow IBE()$ algorithm) and encrypts the measurement; encrypted data is pictured as a vault.
\M\ also tags the report using the secret acquired offline
and a list of keywords characterizing the report; in the example \M\ uses keywords ``Temp'' and ``Irvine, CA''.
Our tagging mechanism leverages the properties of bilinear maps to make sure that, if \M\ and \Q\ use the same keywords, they will compute the same tag,
despite each of them is using a different secret (\M\ is using the grey key while \Q\ is using the yellow one).
As before, the tag and the encrypted report leak no information about the nature of the report or the nominal value of the measurement.
Both tag and encrypted data are forwarded to the Service Provider.

\descr{Report Delivery.}
The Service Provider only needs to match tags sent by Mobile Nodes with the ones uploaded by Queriers.
If the tags match, the corresponding encrypted report is forwarded to the Querier.
In the example of Figure \ref{fig:operations} the green tag matches the blue one, so the encrypted report (the vault)
is forwarded to \Q.
Finally, \Q\ can decrypt the report using the decryption key and recover the temperature measurement.

\subsection{PEPSI overhead}
Resources in \PS\ are not as constrained as in WSNs,
nonetheless, overhead incurred at Mobile Nodes should still be minimized.
To foster the adoption of our solution in current \PS\ applications we provide an experimental evaluation
of the cost of cryptographic operations used to achieve intended privacy features.
We implemented protocol operations executed by Mobile Nodes
on a Nokia N900 (equipped with a 600 MHz ARM processor and 256 MB RAM).
Computation overhead, for every report, is due to the computation of the tag
and the encryption of the measurement.
In our experiments, we experience an average time (over $100$ trials) of $93.47 ms$ to perform
these operations.

Communication overhead is merely due to the transmission of the tag, which is the output of a
hash function (e.g., SHA-1), thus, it is relative small ($160$-bit).
The encryption of the measurement generates almost no overhead, since,
using state-of-the-art symmetric-key ciphers (e.g., AES), ciphertext's length
is almost the same as plaintext's.

Tag computation by Queriers is performed only once, during query subscription.
Upon reception of measurement of interests, Queriers perform symmetric-key decryption,
which incurs a negligible overhead.

Finally, note that the Service Provider incurs no communication nor computational overhead:
its task is limited to comparing output of hash functions (i.e., tags) and forwarding reports.
From a functional point of view, the work of the Service Provider is no different from that in a non privacy-preserving solution.
Thus, privacy protection incurs no overhead at the Service Provider and enjoys scalability to large-scale scenarios.
We conclude that our architecture is practical enough, today, to be deployed for real-world \PS\ applications.

\section{Related Work}\label{sec:related}


\descr{Participatory Sensing Projects.}
In the last few years, Participatory Sensing initiatives have multiplied, ranging from research prototypes
to deployed systems.
Due to space limitations we briefly review some \PS\ application that apparently expose participant privacy (e.g., location, habits, etc.).
Each of them can be easily enhanced with our privacy-protecting layer.
Interested readers may find a larger list of \PS\ applications at \cite{projectpage}.
Quake-Catcher \cite{quakecatcher} aims at building the world's largest,
low-cost strong-motion seismic network by utilizing accelerometers embedded in any internet-connected device.
Kim et al. \cite{kim09ubicomp} use the power of \PS\ for meaningful places (e.g., home, office, etc.) discovery.
\PS\ has been shown to be an effective mean to monitor levels of air pollution \cite{paulos07sensys}, noise pollution \cite{maisonneuve09ITEE} and water quality \cite{kuznetsov10dis}.
\PS\ to aid health care providers in patient monitoring has been investigated in \cite{health-estrin}.

\descr{Privacy.}
Only little attention has been paid to arising privacy issues in \PS\ \cite{shilton}.
The authors of \cite{mobisys08} study privacy in participatory sensing relying on
weak assumptions: they attempted to protect {\em anonymity} of Mobile Nodes
through the use of Mix Networks.
(A Mix Network is a statistical-based anonymizing infrastructure
that provides $k$-anonymity -- i.e., an adversary cannot tell a user from a set of $k$).
However, Mix Networks are unsuitable for many \PS\ settings.
They do not attain provable privacy guarantees and assume the presence of an ubiquitous WiFi infrastructure used by Mobile Nodes,
%
whereas, \PS\ applications do leverage the increasing use of broadband 3G/4G connectivity.
In fact, an ubiquitous presence of open WiFi networks is not realistic today nor
anticipated in the next future.
By contrast, our work aims at identifying a minimal set of realistic assumptions 
and clear privacy guarantees to be achieved with provable security.

The work in  \cite{infocom10} studies privacy-preserving data aggregation,
(e.g., computation of sum, average, variance, etc.). Similarly,~\cite{ganti2008} presents a solution for
community statistics on time-series data, while protecting anonymity (using data perturbation in a closed community with a
known empirical data distribution). Finally,~\cite{gilbert2010} aims at guaranteeing integrity and authenticity of user-generated contents, by employing Trusted Platform Modules (TPMs).

The main technical challenge in providing provable privacy in participatory sensing infrastructure
stems from the simultaneous presence of several mutually untrusted (and potentially unknown) entities, including data producers,
data consumers, and Service Providers. A similar scenario arises in the context of {\em Publish-Subscribe}
networks~\cite{pubsub}, which face similar privacy concerns.
However, state-of-the-art solutions (e.g., \cite{ion10securecomm}) assume an a-priori knowledge (and key exchange) between publishers and subscribers,
while \PS\ application require loose coupling between Mobile Nodes and Queriers.
This makes impossible to apply them to the \PS\ scenario, where data producers and consumers may not know each other.
Our solution protects their privacy while requiring no direct interaction between the two parties.

\section{Conclusion \& Open Problems}

Participatory Sensing is a novel computing paradigm that bears a great potential.
If users are incentivized to contribute personal device resources, a number of novel applications and
business models will arose.
In this article we discussed the problem of protecting privacy in Participatory Sensing.
We claim that user participation cannot be afforded without protecting the privacy
of both data consumers and data producers.
We also proposed the architecture of a privacy-preserving Participatory Sensing infrastructure and
introduced an efficient cryptographic solution that achieves privacy with provable security.
Our solution can be adopted by current Participatory Sensing applications to enforce privacy and
enhance user participation, with little overhead.

This work represents an initial foray into robust privacy guarantees in \PS,
thus, much remains to be done. 
%
Items for future work, include (but are not limited to):
\begin{enumerate}
\item 
Protecting query privacy with respect to
the Registration Authority. Recall, in fact, that Querier Alice needs to obtain the IBE decryption keys
from the Registration Authority, which would then learn Alice's query interests.
\item Protecting node privacy with respect to the Network Operator.
Current technology does not allow to hide users' locations and identities from to the Network Operator.
Hence, it is an interesting challenge to guarantee node anonymity in broadband networks.
\item Addressing collusion attacks, where multiple entities might collaborate in order
to violate the privacy of Mobile Nodes or Queriers. 
%
\item 
Improving the syntax of supported query types. 
In fact, PEPSI so far allows query/report matching based on the tags provided by both Mobile Nodes and Queriers. However,
\PS\ applications might require more complex queries where Queriers are interested in an aggregate of the
reports (e.g., average or sum), 
or even complex query predicates (e.g., comparisons). 
While simple aggregate function evaluation over encrypted data is viable with available cryptographic techniques (e.g., homomorphic encryption), enabling {\em efficient} evaluation of complex predicates remains an open challenge.

\end{enumerate}

\bibliographystyle{abbrv}

\pagebreak

%
%

\end{document}